\documentstyle[preprint,aps,epsfig,amssymb]{revtex}

\begin{document}

\draft
\title{Superconductivity in ferromagnetic metals and
in compounds without inversion centre} 
\author{V.P. Mineev } 
\address{
Commissariat \`{a} l'Energie Atomique, DSM/DRFMC/SPSMS\\
17 rue des Martyrs, 38054 Grenoble Cedex 9, France}
\date{\today}

\maketitle

\begin{abstract}
The symmetry properties and the general overview of the
superconductivity theory in the itinerant ferromagnets and in materials
without space parity are presented.  The basic notions of unconventional
superconductivity are introduced in broad context of multiband
superconductivity which is inherent property of ferromanetic metals or
metals without centre of inversion.

\end{abstract}

\section{Introduction}

The recent discoveries of several materials $UGe_{2}¥$, \cite{1,2} 
$ZrZn_{2}¥$,\cite{3} and $URhGe$ \cite{4,5}, where the superconductivity
coexists with presumably itinerant ferromagnetism, put forward the
problem of theoretical description of such type of ordered media. 
Along with the problem of a mechanism of pairing and critical temperature
calculation which is intensively discussed now in literature in frame of
different models \cite{6,7,8,9,10,11,12,13,14,15,16,17,18}, but
still being far from its resolution, there was developed also the
general symmetry approach to the theoretical description of
superconductivity in ferromagnetic superconductors.  The
superconducting state in these materials have to be preferably spin
triplet to avoid the large depairing influence of the exchange field. 
However, the theory of triplet superconductivity in ferromagnets
cannot be simple replica of the theory of superfluid phases of liquid
$^{3}¥He$ \cite{19} or even its anisotropic strong spin-orbit coupling
generalization for superconducting states in crystals
\cite{20,21,22,23,24}.  Here the superconducting states appear from
another ordered state - namely,  ferromagnetic state.  The latter has
the broken time reversal symmetry and the classifcation of the
superconducting states has its own specifics \cite{25,26,27,28}.

More or less at the
same time another important achievement in the physics of
superconductivity is committed.  This time it was related with
discovery of $MgB_{2}¥$ - the first superconducting material with two
bands of conduction electrons where the existance of two energy gaps
has been unambiguously demonstrated by thermodynamic and spectroscopic
measurements \cite{29}.  
Theoretical investigations of two-band superconductivity have been
undertaken soon after BCS theory \cite{30} were also restarted (see for
example \cite{31,32}).  Recently it was revealed that the
superconductivity in itinerant ferromagnetic superconductors and in
conventional two band superconductors has a lot of similarity
\cite{33}.  Indeed, in ferromagnetic superconductors the different
bands filled by spin "up" and spin "down" electrons are always
present.  Hence one can construct the theoretical description of such
the superconducting states in analogy with conventional multiband
superconductivity.

Also, quite recently the first unconventional superconductor without inversion
symmetry $CePt_{3}¥Si$ has been discovered \cite{34}.
The microscopic
theory of superconductivity in metals without inversion has been
developed by V.Edel'stein \cite{35} pretty long ago.
The different
aspects of theory of superconductivity in such type materials has been
discussed about the same time \cite{36,37,38} and has been advanced
further in more recent publications \cite{39,40,41,42,43,44,45,46}. 
Finally, the general symmetry approch to the superconductivity in the
materials with space parity violation has been developed 
\cite{47,48} and some its applications have been considered
\cite{49,50,51}.  It proves and we shall demonstrate it clearly below
that again the description of superconductivity in such type of
materials has many common features with conventional superconductivity
in two band superconductors.

So, having in mind to present here the review of symmetry approach to the
superconductivity in ferromagnetic materials and in the compounds
without inversion centrum (ferroelectrics) we shall discuss the
normal state properties of the materials with different symmetry. 
There will be shown how the symmetry of normal metal one-electron
states determines the possible types of pairing.  We shall follow as
far as it possible to analogy with multiband conventional
superconductivity.  We make an overview of symmetry description of
superconducting states specific for materials with time reversal
or space parity violation.  At last the Gor'kov type formalism for multiband
superconductivity in ferromagnetic and ferroelectric materials will be
developed and some concrete applications will be discussed.

\section{One band superconductivity}

To introduce notations and basic notions
we start with well known description of the superconductivity in one
band normal metal with centrum of space inversion \cite{20,21,22,23,24}.  The band
states are Bloch type wave functions characterized by quasimomentum
${\bf k}$ and $\pm 1/2$ projections of spin on the direction of the quantization
axis.  Each electronic eigen state is fourth fold degenerate, another
words, four different states $|{\bf k},\uparrow\rangle $, $|-{\bf
k},\uparrow\rangle $, $|{\bf k},\downarrow\rangle $ and $|-{\bf
k},\downarrow\rangle$ correspond to the same electron energy
$\varepsilon_{{\bf k}}¥$.  The states with opposite momenta and the
opposite spins form so called Kramers doublet: under the time
inversion operation $\hat K$ they are transformed to each other with a
phase factor accuracy $\hat K|{\bf k}, \uparrow\rangle=e^{i\phi}¥
|-{\bf k},\downarrow\rangle$.  In its own turn the states with the
opposite momenta are transformed to each other by means of the space
parity operation $\hat I$.  So, the presence of four degenerate
electronic states $|{\bf k},\uparrow\rangle$, $\hat K|{\bf
k},\uparrow\rangle$, $\hat I|{\bf k},\uparrow\rangle$, and $\hat K\hat
I|{\bf k},\uparrow\rangle$ is the consequence of space and time
inversion symmetries.

Now, if there is some pairing interaction, the scattering of
electrons occupying the degenerate states with oppositely directed
momenta near the Fermi surface and either antiparallel spins $(S=0)$ or
parallel spins $(S=1)$ results in formation of superconducting state
with order parameters depending of space coordinate ${\bf r}$ and the
relative direction of momenta of pairing particles ${\hat{\bf k}}=({\bf
k}_{1}¥-{\bf k}_{2}¥)/2k_{F}¥$:
\begin{equation}
\Delta_{\alpha \beta}^{S=0}¥({\bf r},\hat{\bf k})= 
\Delta({\bf r},\hat {\bf k})i\sigma_{y}¥= \Delta({\bf r},\hat {\bf k})
(|\uparrow\downarrow\rangle-|\downarrow\uparrow\rangle),
\label{e1}
\end{equation}
\begin{eqnarray}
\Delta_{\alpha \beta}^{S=1}¥({\bf r},\hat {\bf k})&=& \left(
\begin{array}{cc}
\Delta_{\uparrow}¥({\bf r},\hat {\bf k}) & \Delta_{0}¥({\bf r},\hat {\bf k}) \\
\Delta_{0}¥({\bf r},\hat {\bf k}) &\Delta_{\downarrow}¥({\bf r},\hat {\bf k})
\end{array}\right)=\Delta_{\uparrow}¥({\bf r},\hat{\bf
k})|\uparrow\uparrow\rangle +\Delta_{0}¥({\bf r},\hat{\bf
k})(|\uparrow\downarrow\rangle+|\downarrow\uparrow\rangle)
+\Delta_{\downarrow}¥({\bf r},\hat{\bf
k})|\downarrow\downarrow\rangle\nonumber\\
&=&\left({\bf d}({\bf r},{\bf k}) \bbox{\sigma}\right)i\sigma_{y}
=\left(
\begin{array}{cc}
-d_{x}({\bf r},\hat {\bf k})+i d_{y}({\bf r},\hat {\bf k})&
d_{z}({\bf r},\hat{\bf k}) \\
d_{z}({\bf r},\hat {\bf k}) & d_{x}({\bf r},\hat {\bf k})+
i d_{y}({\bf r},\hat {\bf k}),
\end{array}
\right),
\label{e2}
\end{eqnarray}
where  $\bbox{\sigma}=(\sigma_{x},\sigma_{y},\sigma_{z})$ are the 
Pauli matrices.

According the Pauli principle the particle permutation , that is
interchange by places of first and second arrow in each
$|\ldots\rangle$ and change $\hat{\bf k}\to -\hat{\bf k} $, yields the
change of sign of the order parameter (pair wave function).  That is
why the singlet pairing states are described only by even $\Delta(-\hat{\bf
k})=\Delta(\hat{\bf k})$ functions whereas triplet pairing states are always
odd $\Delta_{\uparrow, \downarrow,0}¥(-\hat{\bf k})=-
\Delta_{\uparrow, \downarrow,0}¥(\hat{\bf k})$ in respect to relative
wave vector direction of pairing particles.  In the crystal with space
inversion the parity of the superconducting state has definite value
and the mixture of the singlet and triplet pairing states
inadmissible.

The scalar function of the order parameter for the siglet
superconducting state is decomposed over the functions 
$\psi_{i}¥(\hat{\bf k})$ of irreducible
representation $\Gamma$ dimensionality $d$ of the point symmetry group
$G$ of the crystal in the normal state
\begin{equation}
\Delta({\bf r},\hat{\bf k})=
\sum_{i=1}^d\eta_{i}¥({\bf r})\psi_{i}¥(\hat{\bf k}).
\label{e3}
\end{equation}
Similar decomposition takes place for vectorial order parameter
function in triplet state
\begin{equation}
{\bf d}({\bf r},{\bf k})= \sum_{i=1}^d\eta_{i}¥({\bf
r})\bbox{\psi}_{i}¥(\hat{\bf k}).
\label{e4}
\end{equation}
Here 
$$\bbox{\psi}_{i}¥(\hat{\bf k})=\hat x{\psi}_{xi}¥(\hat{\bf k})+
\hat y{\psi}_{yi}¥(\hat{\bf k})+\hat z{\psi}_{zi}¥(\hat{\bf k})$$
are vectorial functions  of irreducible
representation $\Gamma$ dimensionality $d$ of the point symmetry group
$G$ represented as decomposition over spin unit vectors
$\hat x$, $\hat y$, $\hat z$ pinned to the crystal axis.  To each irreducible
representation corresponds its own critical temperature $T_{c}¥$.  Any
one-dimensional representation describes only one superconducting
state characterized by its own symmetry group - so called
{\bf superconducting class} which is a subgroup of the group of symmetry of
the normal state $G\times I\times U(1)\times K$, where $U(1)$ is the
group of gauge transformations.  A multidimensional representation
gives rise to several superconducting states.  Their order parameters
are given by the particular linear combinations in (\ref{e3}) or in
(\ref{e4}) possessing of different symmetries, another words, belonging to
different superconducting classes but being characterized by the same
critical temperature.

\section{Multi band superconductivity}

Let us look now what kind of modifications must be introduced in the theory
if we deal with superconducting state forming in a metal with several
conduction bands.  We shall  speak for simplicity about two band situation. 
Each band has its own dispersion law $\varepsilon_{\lambda}¥({\bf
k})$, here $\lambda$ is the band index, and its own Fermi surface
determined by equations $\varepsilon_{1}¥({\bf k})=\varepsilon_{F}¥$
and $\varepsilon_{2}¥({\bf k})=\varepsilon_{F}¥$.  The electronic
states in each of two bands obey the same fourfold degeneracy as
before.

If there
is some pairing interaction it acts in some energetic vicinity of
Fermi surface.  One can also say that the pairing of electronic states
in given band happens in some layer in the reciprocal space around
corresponding Fermi surface.  The different bands layers where the
pairing takes place can in principle intersect each other.  Then the
pairing is possible not only between the electrons occupying the
electronic states from  the same band but also between the electrons
from the different bands with equal by the modulus and oppositely
directed momenta and either antiparallel or parallel spins.  On
microscopic level of description we have to introduce not only doubled
set of Bogolubov or Gor'kov equations written for each band but also
additional equations for the Bogolubov interband pairing amplitudes or
for the Gor'kov interband $F_{12}¥$ Green functions.  Hence the total
system includes 6 coupled equations.  On macroscopic level of the
description  we shall have
coupled system of Ginzburg-Landau equations which, even in the
simplest case of superconducting state corresponding to
one-dimensional representation, will consists of three equations for
three coordinate dependent pairing amplitudes.  These amplitudes are
different complex functions with the coinciding phase factors (see
below).  In the linear approximation the system of Ginzburg-Landau
equations has tree different eigen values the largest of which
determines the critical temperature or the upper critical field for
the phase transition to the superconducting state.  So, the multiband
superconductivity has the pecularities in the mathematical
description.

Let us consider now more simple situation when the different bands
layers in reciprocal space, where the pairing takes place, do not intersect each
other.  Then the pairing is possible only between the electrons with
the opposite momenta ${\bf k}$ and $-{\bf k}$ occupying electronic states in
the same band.  Certainly, in principle, one can consider also interband
pairing.  But the Cooper pairs formed by electrons from the different bands 
with different Fermi momenta $k_{F1}¥(\hat {\bf k})$ and $k_{F2}¥(-\hat {\bf k})$
will inevitably
have a finite momentum and as result one can expect of appearance only
of space modulated or Fulde-Ferrell-Larkin-Ovchinnikov superconducting
state.  We shall not discuss here this exotic possibility.

So, if we consider only intraband pairing we  deal with two sets of
Gor'kov equations written for each band.  The coupling between them is
carried out only due to the processes interband pair transitions which
is taken into account by means of the self-consistency equation.  Thus,
in general, the order parameter consists of two parts of the same form
as for one-band superconductivity.  For the siglet superconducting
state instead (\ref{e3}) we have
\begin{equation}
\Delta({\bf r},\hat{\bf k})=
\sum_{\lambda=1,2}¥\sum_{i=1}^d\eta_{i\lambda}¥({\bf
r})\psi_{i\lambda}¥(\hat{\bf k}),
\label{e5}
\end{equation}
where $\psi_{i1}¥(\hat{\bf k})$ and $\psi_{i2}¥(\hat{\bf k})$  
are in principle different functions of the same irreducible
representation $\Gamma$ dimensionality $d$ of the point symmetry group
$G$ of the crystal in the normal state.  Similar decomposition takes
place for vectorial order parameter function in triplet state
\begin{equation}
{\bf d}({\bf r},{\bf k})= \sum_{\lambda=1,2}\sum_{i=1}^d\eta_{i\lambda}¥({\bf
r})\bbox{\psi}_{i\lambda}¥(\hat{\bf k})
\label{e6}
\end{equation}

Let us take for clarity the case of one-dimensional representation,
for instance conventional singlet two-band superconductivity which is
intensively discussed now in connection with $MgB_{2}¥$ compound.  In this
case the order parameter has the form
\begin{equation}
\Delta({\bf r},\hat{\bf k})=
\eta_{1}¥({\bf r})\psi_{1}¥(\hat{\bf k})+\eta_{2}¥({\bf r})\psi_{2}¥(\hat{\bf k}),
\label{e7}
\end{equation}
where
the coordinate dependent complex order parameter amplitudes
$\eta_{1}¥({\bf r})$ and $\eta_{2}¥({\bf r})$
are not completely independent:
\begin{equation}
\eta_{1}¥({\bf r})=|\eta_{1}¥({\bf r})|e^{i\varphi({\bf r})}¥,
~~~\eta_{2}¥({\bf r})=\pm|\eta_{2}¥({\bf r})| e^{i\varphi({\bf r})}.
\label{e8}
\end{equation}
Thus, being different by their modulos they have the same phase with an
accuracy $\pm \pi$.  The latter property guarantees the consistency of
transformation of both parts of the order parameter under the time
reversal.

In the space homogeneous case the coupled system of Ginzburg-Landau
equations has the form
\begin{eqnarray}
\eta_{1}¥ &=&(g_{1}¥\eta_{1}¥+
g_{12}¥\eta_{2}¥)\lambda(T_{c0}¥),\nonumber\\
\eta_{2}¥&=&(g_{21}¥\eta_{1}¥+
g_{2}¥\eta_{2}¥)\lambda(T_{c0}¥).
\label{e9}
\end{eqnarray}
The function $\lambda(T)$ is 
\begin{equation}
\lambda(T)=2\pi T\sum_{n\ge 0}¥\frac{1}{\omega_{n}¥}=\ln
\frac{2\gamma \epsilon}{\pi T}~,
\label{e10}
\end{equation}
$\ln\gamma=0,577\ldots $ is the Euler constant, $\epsilon $ is an
energy cutoff.
 
Thus, 
the critical temperature is given by \cite{30,31} 
\begin{equation}
T_{c0}¥=(2\gamma\epsilon/\pi)\exp{(-1/g)},
\label{e11}
\end{equation}
where $g$ is defined by the maximum of zeros of determinant of the system
(\ref{e9})
\begin{equation}
g= (g_{1}¥+g_{2}¥)/{2} +\sqrt{({g_{1}¥-g_{2}¥})^{2}¥/{4}+g_{12}¥g_{21}¥}.
\label{e13}
\end{equation}
In particular at $g_{12}¥,~g_{21}¥~\ll~g_{1}¥,~g_{2}¥$ the critical
temperature is determined by
\begin{equation}
g= \max(g_{1}¥,~g_{2}¥).
\label{e14}
\end{equation}

\section{Ferromagnetic superconductors with triplet pairing}

\subsection{ The order parameters}

In an itinerant feromagnetic metal
the internal exchange field lifts the Kramers degeneracy of the
electronic states.  The electrons with spin "up" fill the states in
one band and the electrons with spin "down" occupate the states in
another band.  Hence we have the specific example of multiband metal
with states in each band filled by electrons with only one spin
direction.  This situation has relation also to the ferromagnetic
metal where ferromagnetic moment originates from almost localized
$f$-shells whereas for its metallic properties other electronic states
from $s$ and $d$ delocalized bands are responsible.  The exchange
interaction is provided here for instance by Ruderman-Kittel
mechanism.  At the same time the back influence of ordered magnetic
moments on the conduction electrons splits the bands with spin up and
spin down electronic states magnetizing the conduction electron
system.

Let us discuss for simplicity the two-band ferromagnet. 
Again as in the case of two band normal metal, if there is some
pairing interaction, one can discuss intraband or spin "up" - spin
"up" ( spin "down"-spin "down") pairing of electrons, as well as
interband or spin "up"-spin "down" pairing.  In general the Fermi surfaces
of spin up and spin down bands are situated in different places of
the reciprocal space and have the different shape.  That is why
pairing of electrons from the different bands occurs just in the case
of nesting of some peaces of the corresponding Fermi surfaces.  In
such the situation, similar to SDW or CDW ordering, the
superconducting ordering is formed by Cooper pairs condensate with
finite momentum known as Fulde-Ferrel-Larkin-Ovchinnikov state.  We
shall not discuss here this special possibility.  So we neglect by
pairing of electronic states from different bands giving Cooper pairs
with zero spin projection.  Hence, the only superconducting state
should be considered it is the state with triplet pairing and the
order parameter given by
\begin{equation}
{\bf d}^{\Gamma}({\bf R},{\bf k})=\frac{1}{2}
[-(\hat{x}+i\hat{y})\Delta_{\uparrow}¥({\bf R},{\bf k})+
(\hat{x}-i\hat{y})\Delta_{\downarrow}¥({\bf R},{\bf k})]
\label{e14}
\end{equation}
Superconducting states ${\bf d}^{\Gamma}({\bf 
R},{\bf k})$ with different critical temperatures in the
ferromagnetic crystals are classified in accordance with irreducible
co-representations $\Gamma$ of the magnetic group $M$ of
crystal~\cite{26,28}.  All the co-representations in ferromagnets with
orthorombic, hexagonal, and cubic symmetries are one-dimensional. 
However, they
obey of multicomponent order parameters determined through the
coordinate dependent pairing amplitudes: one per each band populated
by electrons with spins "up" or "down".  For the two-band
ferromagnet under discussion, they are
\begin{equation} 
\Delta_{\uparrow}¥({\bf R},{\bf k})=-\eta_{1}¥({\bf R})f_{-}¥({\bf
k}),~~~~ \Delta_{\downarrow}¥({\bf R},{\bf k})=\eta_{2}¥({\bf
R})f_{+}¥({\bf k}).
\label{e15}
\end{equation}
The coordinate dependent complex order parameter amplitudes
$\eta_{1}¥({\bf R})$ and $\eta_{2}¥({\bf R})$ 
are not completely independent:
\begin{equation}
\eta_{1}¥({\bf R})=|\eta_{1}¥({\bf R})|e^{i\varphi({\bf R})}¥,
~~~\eta_{2}¥({\bf R})=\pm |\eta_{2}¥({\bf R})| e^{i\varphi({\bf R})}.
\label{e16}
\end{equation}
As in conventional two-band case (\ref{e8}), being different by their modulos they
have the same phase with an accuracy $\pm \pi$.  The latter property
is due to the consistency of transformation of both parts of the order
parameter under the time reversal.

The general forms of odd functions of momentum directions of pairing
particles on the Fermi
surface $f_{\pm}¥({\bf k})=f_{x}¥({\bf k})\pm if_{y}¥({\bf k})$ for the
different superconducting states in ferromagnets can be found
following the procedure introduced in \cite{28} and it is described
here on example of ferromagnetic orthorombic crystal.  The same
results for the ferromagnets with orthorombic and cubic symmetry 
in terms of the functions  
$f_{x}¥({\bf k})$ and $f_{y}¥({\bf k})$ one can find in the paper \cite{28}.

Let us consider a ferromagnetic orthorombic crystal with spontaneous
magnetization along one of the symmetry axis of the second order
chosen as the $z$-direction.  Its symmetry group
\begin{equation}
    G=M \times U(1)
\label{17} 
\end{equation}
consists of the so called magnetic class \cite{52}, or black-white group 
$M$ having a white subgroup $H$ of index 2,
and the group of the gauge transformations $U(1)$.  In the given
case $M$  is equal to $D_{2}(C_{2}^{z})=(E,
C_{2}^{z})+ KC_{2}^{x}\times (E, C_{2}^{z}) =(E, C_{2}^{z}, KC_{2}^{x},
KC_{2}^{y})$, where $K$ is the time reversal operation, and $H=(E,
C_{2}^{z})$.  The symmetry of any magnetic superconducting state
arising directly from this normal state corresponds to the one of the
subgroups of the group $G$ characterized by broken gauge symmetry.  As
it was already mentioned the superconducting states ${\bf
d}^{\Gamma}({\bf R},{\bf k})$ with different critical temperatures in
the ferromagnetic crystals are classified in accordance with
irreducible co-representations $\Gamma$ of the magnetic group $M$ of
crystal.  The irreducible corepresentations of $M$ are derived from
the irreducible representations of $H$.  The whole procedure was
introduced by E.Wigner and well described in \cite{53,54}.  For us,
however, there will be convinient not to follow this general formalism
but discuss first the symmetries of possible superconducting states in
ferromagnetic material, another words, the {\bf ferromagnetic
superconducting classes}.  According to general rules \cite{20} they
have to be given by the subgroups of $G$ constructed by means of
combining elements of $M$ with phase factor $e^{i\pi}¥$ being an
element of the group of the gauge transformations $U(1)$.  The
explicit structures of these subgroups isomorphic to the initial
magnetic group $D_2(C_2^{z}¥)$ are
\begin{equation}
D_{2}(C_{2}^{z})=
(E, C_{2}^{z}, KC_{2}^{x}, KC_{2}^{y}),
\label{e18}
\end{equation}

\begin{equation}
\tilde D_{2}(C_{2}^{z})= (E, C_{2}^{z},
KC_{2}^{x}e^{i\pi},KC_{2}^{y}e^{i\pi}),
\label{e19}
\end{equation}

\begin{equation}
D_{2}(E)= (E, C_{2}^{z}e^{i\pi}, KC_{2}^{x}e^{i\pi}, KC_{2}^{y}),
\label{e20}
\end{equation}
\begin{equation}
\tilde D_{2}(E)= (E, C_{2}^{z}e^{i\pi}, KC_{2}^{x}, KC_{2}^{y}e^{i\pi}).
\label{e21}
\end{equation}

The general forms of the order parameters 
\begin{equation}
{\bf d}^{\Gamma}({\bf R},{\bf k})=\frac{1}{2}
[\eta_{1}¥(\hat{x}+i\hat{y})f_{-}¥^{\Gamma}¥({\bf k}) +
\eta_{2}¥(\hat{x}-i\hat{y})f_{+}¥^{\Gamma}¥({\bf k}) ]
\label{e22}
\end{equation}
compatible with symmetries (\ref{e18})-(\ref{e21}) correspondingly
are obtained by the following choice of the functions $f_{\pm}¥({\bf k})$:
\begin{equation}
f_{\pm}^{A_{1}¥}({\bf k})=
k_{x}¥(u_{1}¥^{A_{1}¥}¥\mp u_{4}¥^{A_{1}¥}¥)
+ik_{y}¥(u_{2}¥^{A_{1}¥}¥\pm u_{3}¥^{A_{1}¥}¥),
\label{e23}
\end{equation}
\begin{equation}
f_{\pm}^{A_{2}¥}({\bf k})=
ik_{x}¥(u_{1}¥^{A_{2}¥}¥\pm u_{4}¥^{A_{2}¥}¥)
+k_{y}¥(u_{2}¥^{A_{2}¥}¥\mp u_{3}¥^{A_{2}¥}¥) ,
\label{e24}
\end{equation}
\begin{equation}
f_{\pm}^{B_{1}¥}({\bf k})=
k_{z}¥(u_{1}¥^{B_{1}¥}¥\mp  u_{4}¥^{B_{1}¥}¥)
+ik_{x}¥k_{y}¥k_{z}¥(u_{2}¥^{B_{1}¥}¥\pm  u_{3}¥^{B_{1}¥}¥) ,
\label{e25}
\end{equation}
\begin{equation}
f_{\pm}^{B_{2}¥}({\bf k})=
ik_{z}¥(u_{1}¥^{B_{2}¥}¥\pm u_{4}¥^{B_{2}¥}¥)
+k_{x}¥k_{y}¥k_{z}¥(u_{2}¥^{B_{2}¥}¥\mp u_{3}¥^{B_{2}¥}¥) ,
\label{e26}
\end{equation}
where $u_{1}¥^{A_{1}¥}¥,\ldots $ are real functions of $k_{x}¥^{2}¥, 
k_{y}¥^{2}¥, k_{z}¥^{2}¥$.

From the expressions for the order parameters one can conlude  that the
only {\bf symmetry dictated nodes in quasiparticle spectrum} of
superconducting A-states in orthorombic ferromagnets are the nodes
lying on the nothern and southern poles of the Fermi surface
$k_{x}¥=k_{y}¥=0$.  On the contrary for the B-states they are on the
line of equator $k_{z}¥=0$.

Similarly, it follows from the general forms of the functions $f_{\pm}({\bf
k})=f_{x}({\bf k})\pm if_{y}¥({\bf k})$ found in \cite{28} that
for $A$ and $B$
superconducting state of tetragonal magnetic class in cubic crystal
the only symmetry nodes are on the nothern and southern poles of the
Fermi surface $k_{x}¥=k_{y}¥=0$ and for for $E_{+}¥$ and $E_{-}¥$
superconducting states they are both on the poles and on the equator line.
For trigonal magnetic class, states $A$ have the nodes on the poles 
(the direction of the polar axis coincides here with the space
diagonal of cube) and states $E_{+}¥$ and $E_{-}¥$ have no symmetry
nodes at all.

The classification of the states in quantum mechanics corresponds
to the general statement by E.Wigner that the different eigenvalues
are related to the sets of eigenstates belonging to the different
irreducible representations of the group of symmetry of the
hamiltonian.  In particular, in absence of the time inversion symmetry
violation, the superconducting states relating to the nonequivalent
irreducible representations of the point symmetry group of crystal
obey the different critical temperatures.  Similarly the eigenstates
of the particles in the ferromagnetic crystals are classified in
accordance with {\bf corepresentations} $\Gamma$ of magnetic group $M$
of the crystal \cite{53,54}.  The latter differs from usual
representations by the {\bf law of multiplication} of matrices of
representation which is
$\Gamma(g_{1}¥)\Gamma(g_{2}¥)=\Gamma(g_{1}¥g_{2}¥)$ for elements
$g_{1}¥, g_{2}¥$ of group $M$ if element $g_{1}¥$ does not include the
time inversion operation and $\Gamma
(g_{1}¥)\Gamma^{*}¥(g_{2}¥)=\Gamma(g_{1}¥g_{2}¥)$ if element $g_{1}¥$
does include the time inversion.  The matrices of transformation of
the order parameters (\ref{e22}) by the symmetry operations of the
group $D_{2}(C_{2}^{z})=(E, C_{2}^{z}, KC_{2}^{x}, KC_{2}^{y})$ are
just numbers (characters).  As usual for one-dimensional
representations they are equal $\pm 1$.  For the state $A_{1}¥$
(\ref{e23}) which is a conventional superconducting state obeying the
complete point-magnetric symmetry of initial normal state they are
$(1,1,1,1)$.  For the order parameter $A_{2}¥$ (\ref{e24}) they are
$(1,1,-1,-1)$ where $-1$ corresponds to the elements of the
superconducting symmetry class (\ref{e19}) containing the phase factor
$e^{i\pi}¥$.  The same is true for the table of characters of the
other states.  So all the corepresentations in the present case are
real, however their difference from the usual representations
manifests itself in the relationship of equivalence.

The two {\bf corepresentations} of the group $M$ are called {\bf equivalent}
\cite{54} if their matrices $\Gamma(g)$ and $\Gamma'(g)$ are
transformed to each other by means of the unitary matrix $U$ as
$\Gamma'(g)=U^{-1}¥\Gamma(g)U$ if the element $g$ does not include the
time inversion and as $\Gamma'(g)=U^{-1}¥\Gamma(g)U^{*}¥$ if the
element $g$ includes the time inversion.  The corepresentations for
the pair of states $A_{1}¥$ and $A_{2}¥$ are equivalent .  In view of
one-dimensional character of these corepresentations the matrix of the
unitary transformation is simply given by the number $U=i$.  The
states $A_{1}¥$ and $A_{2}¥$ belong to the same corepresentation and
represent two particular forms of the same superconducting state.  The
state $\Psi^{A_{2}} $ transforms as $i{\Psi^{A_{1}¥*}}$ and the state
$\Psi^{B_{2}}$ transforms as $i{\Psi^{B_{1}*}}$.  It will be shown
below that if we have state $A_{1}¥$ in the ferromagnet domains with
the magnetization directed up the superconducting state in the domains
with down direction of the magnetization corresponds to the
superconducting state $A_{2}¥$.  The same is true for the pair of
states $B_{1}¥$ and $B_{2}¥$.

The critical temperatures of equivalent states $A_{1}¥$ and $A_{2}¥$,
or equivalent states $B_{1}¥$ and $B_{2}¥$, in the ferromagnetic domains
with the opposite orientations of magnetization are equal (see below).
At the same time the states $A_{1}¥$ and $A_{2}¥$ have the different
symmetries, that means they belong formally to different ferromagnetic
superconducting classes.  In this respect the superconducting states in
ferromagnet transforming according to one-dimensinal corepresentations
of magnetic group $M$ reveal sort of similarity on multicomponent
superconducting states transforming according to multidimentional
representation of the point group of paramagnetic state.

All the listed above superconducting phases are in principle
non-unitary and obey the {\bf Cooper pair spin momentum}
\begin{equation} 
{\bf S}=i\langle {\bf d}^{*}\times {\bf d}\rangle=
\frac{\hat z}{2}\langle|\Delta_{\uparrow}¥|^{2}¥-
|\Delta_{\downarrow}¥|^{2}¥\rangle ,
\label{e27}
\end{equation}
and {\bf Cooper pair angular momentum}
\begin{equation} 
{\bf L}=i\langle {\bf d}^{*}_{\alpha}\left({\bf k}\times
\frac{\partial}{\partial {\bf k}} \right) {\bf d}_{\alpha}\rangle=
\frac{i}{2}\langle \Delta_{\uparrow}¥^{*}\left({\bf k}\times
\frac{\partial}{\partial {\bf k}} \right)\Delta_{\uparrow}¥+
\Delta_{\downarrow }¥^{*}\left({\bf k}\times \frac{\partial}{\partial
{\bf k}} \right)\Delta_{\downarrow } \rangle,
\label{e28}
\end{equation}
where the angular brackets denote the averaging over
${\bf k}$ directions.  As the consequence the magnetic moment
of ferromagnet changes at the transition to the ferromagnetic
superconducting state.  We shall calculate this particular
changement below.

\subsection{Gor'kov equations}

The BCS Hamiltonian in two-band ferromagnet with triplet pairing is
\begin{equation}
H=\sum_{{\bf k},{\bf k}',\alpha}\langle {\bf k} |\hat{h}_{\alpha}¥|{\bf
k}'\rangle a^{\dag}_{ {\bf k}\alpha} a_{{\bf k}'\alpha} +\frac{1}{2} 
\sum_{{\bf k},{\bf k}', {\bf q},\alpha,\beta} V_{\alpha
\beta}({\bf k},{\bf k}') a^{\dag}_{
-{\bf k}+{\bf q}/2, \alpha} a^{\dag}_{ {\bf k} +{\bf q}/2,\alpha} a_{
{\bf k}' +{\bf q}/2,\beta} a_{ -{\bf k}'+{\bf q}/2, \beta},
\label{e29}
\end{equation}
where the band indicies $\alpha$ and $\beta$ are $(\uparrow,
\downarrow)$ or $(1,2)$,
\begin{equation}
\hat{h}_{\alpha}¥=\hat{\varepsilon}_{\alpha}¥ 
-\mu_{B}¥\hat{\bf g}_{\alpha}¥{\bf H}/2 +U({\bf
r})-\varepsilon_{F},
\label{e30}
\end{equation}
are one particle band energy operators, the
functions $\hat{\varepsilon}_{\alpha}¥ $ (including the exchange splitting)
and $\hat{\bf g}_{\alpha}¥~$- factor depend of projections of gauge invariant
operator $-i \nabla +({e}/{c}) {\bf A}({\bf r})$ on crystallografic
directions.  In the simplest case of isotropic bands without a
spin-orbital coupling ${\bf g}_{1,2}¥=\pm 2{\bf H}/H$.  $U({\bf r})$
is an impurity potential, ${\bf A}({\bf r})$ is vector potential such that
\begin{equation}
\nabla \times {\bf A} = {\bf B}={\bf H}+4 \pi{\bf M},
\label{e31}
\end{equation} 
${\bf M}$ is the magnetic moment of the ferromagnet, ${\bf H}$ is a
magnetic field, which should be determined from the Maxwell equations
\begin{equation}
\nabla \times {\bf H} =\frac {4\pi}{c}{\bf j}, ~~~\nabla{\bf B}=0,
\label{e32}
\end{equation}
with Maxwell bondary conditions of the continuity of $B_{n}¥$ and
$ H_{t}¥$ at the boundary of the sample and ${\bf H}\to{\bf H}_{ext}¥$
at infinity.   The equations for
determination the moment ${\bf M}$ and current ${\bf j}$ densities see
below.

The pairing potential interaction is expanded over
\begin{equation}
V_{\alpha\beta}({\bf k},{\bf k}') =-V_{\alpha\beta}\varphi_{\alpha}¥
({\bf k})\varphi_{\beta}¥^{*}¥({\bf k'}),
\label{e33}
\end{equation}
where
\begin{equation}
\varphi_{\uparrow}({\bf k})=-f_{-}({\bf k}),~~~~
\varphi_{\downarrow}({\bf k})=f_{+}({\bf k}).
\label{e34}
\end{equation}
It contains four different interaction terms corresponding to: (i) a pairing
between electrons with the same spin polarization (intraband
interaction) and (ii) the interband scattering terms with 
$V_{\uparrow\downarrow}¥=V_{\downarrow\uparrow}¥$ describing the transitions of
the pair electron from one sheet of the Fermi surface to the other
sheet by reversing the pair spin orientation with the help of the
spin-orbit coupling.

When the interband scattering is negligible $V_{\uparrow\downarrow}¥=
V_{\downarrow\uparrow}¥=0$, the pairing of the electrons occurs first
only in one of the sheets of the Fermi surface like in the $A_{1}$
phase of $^{3}$He.  In general the superconductivity in each band is
not independent.

The full system of equations describing the magnetostatic behavior of
ferromagnetic superconductor consists of Gor'kov equations for the
Green functions in two bands, 
\begin{eqnarray}
\sum_{{\bf k}_{1}}&\langle& {\bf k}|i\omega_{n}-\hat h_{\alpha}| {\bf
k}_{1}\rangle G_{\alpha}\left({\bf k}_{1},{\bf k}',\omega_{n}\right)
+\sum_{{\bf q}} \Delta_{\alpha}\left({\bf k},{\bf q}\right)
F^{\dag}_{\alpha}\left({\bf k}-{\bf q},{\bf k}',\omega_{n}\right)
=\delta\left({\bf k}-{\bf k}'\right), \\
\sum_{{\bf k}_{1}}&\langle& -{\bf k}_{1}|i\omega_{n}+\hat h_{\alpha}| -{\bf
k}\rangle F^{\dag}_{\alpha}\left({\bf k}_{1},{\bf k}',\omega_{n}\right)
+\sum_{{\bf q}} \Delta^{\dag}_{\alpha}\left({\bf k},{\bf
q}\right) G_{\alpha}\left({\bf k}+{\bf q},{\bf k}',\omega_{n}\right)
=0, \\
\sum_{{\bf k}_{1}}&\langle &{\bf k}|i\omega_{n}-\hat h_{\alpha}| {\bf
k}_{1}\rangle F_{\alpha}\left({\bf k}_{1},{\bf k}',\omega_{n}\right)
-\sum_{{\bf q}} \Delta_{\alpha }\left({\bf k},{\bf q}\right)
G_{\alpha}\left(-{\bf k}',-{\bf k}+{\bf q},\omega_{n}\right) =0,
\label{e35}
\end{eqnarray}
combined with the self-consistency equation
\begin{eqnarray}
\Delta_{\alpha }\left({\bf k},{\bf q}\right)=-T\sum_{n} \sum_{{\bf
k}'} V_{ \alpha \beta}({\bf k},{\bf k'}) F_{\beta} \left({\bf
k}'+\frac{{\bf q}}{2},{\bf k}'-\frac{{\bf q}}{2},\omega_{n} \right),
\label{e36}
\end{eqnarray}
here $\omega_{n}=(2n+1)\pi T$ are fermionic Matsubara frequencies, 
Maxwell equations (\ref{e31}), (\ref{e32}) and definitions of current density
\begin{equation}
{\bf j}({\bf k})=2{e}T\sum_{n}\sum_{{\bf
p}{\bf
p}'}¥\sum_{\alpha=\uparrow,\downarrow}\left(\frac{\partial}{\partial{\bf p}}+
\frac{\partial}{\partial{\bf p}'}\right)
\langle {\bf p}|\hat \varepsilon_{\alpha}¥|{\bf p}'\rangle G_{\alpha}\left (
\frac{{\bf p }+{\bf p }'+{\bf k}}{2},\frac{{\bf p }+{\bf p }'-{\bf k}}{2},\omega_{n}
\right),
\label{e37}
\end{equation}
and magnetic moment density
\begin{equation}
{\bf M}({\bf k})=\hat z \mu_{B}¥T\sum_{n}\sum_{{\bf
p}}¥[G_{\uparrow}({\bf p}+{\bf k}/2,{\bf p}-{\bf k}/2,\omega_{n})-
G_{\downarrow}({\bf p}+{\bf k}/2,{\bf p}-{\bf k}/2,\omega_{n})].
\label{e38}
\end{equation}

We want to determine  the Green's functions of ferromagnetic 
superconductors in the absence of external field and impurity scattering. 
Even under these simple conditions, the system is not spatially
uniform due to the inherent presence of $ 4\pi M$.  If we artificially
neglect by $4 \pi M$ as internal field acting diamagnetically on
electron charges taking $ {\bf A} =0$, the system is spatially
uniform.  Then, we can write the Gor'kov equations in the form
\begin{eqnarray}
\left(i\omega_{n}-\xi_{{\bf k}\alpha}\right) G_{\alpha}({\bf k},\omega_{n})+
\Delta_{\alpha}({\bf k}) F_{\alpha}^{\dagger}({\bf k},\omega_{n})=1
 \\
\left(i\omega_{n}+\xi_{{\bf k}\alpha}\right) F_{\alpha}^{\dagger}¥ ({\bf
k},\omega_{n})+\Delta_{\alpha}^{\dagger}¥ ({\bf k}) G_{\alpha}({\bf
k},\omega_{n})=0
\label{e39}
\end{eqnarray}
where $\xi_{{\bf k}\alpha}=\varepsilon_{{\bf k}\alpha}-\varepsilon_{F}$.
The equations for each band are only coupled through the order
parameter given by the self-consistency condition
\begin{equation}
\Delta_{\alpha}({\bf k})=-T\sum_{n}\sum_{{\bf
k}'}\sum_{\beta=\uparrow,\downarrow} V_{\alpha,\beta}\left( {\bf k},{\bf
k}'\right) F_{\beta}({\bf k}',\omega_{n}).
\label{e40}
\end{equation}
The superconductor Green's functions are
\begin{eqnarray}
G_{\alpha}\left({\bf k},\omega_{n}\right) &=& -\frac{i\omega_{n}+\xi_{{\bf
k}\alpha}}{\omega_{n}^{2}+E_{{\bf k},\alpha}^{2}} \\
F_{\alpha}\left({\bf k},\omega_{n}\right)&=& \frac{\Delta_{\alpha} ({\bf
k})}{ \omega_{n}^{2}+E_{{\bf k},\alpha}^{2} },
\label{e41}
\end{eqnarray}
where
$E_{{\bf k},\alpha}=\sqrt{\xi_{{\bf
k}\alpha}^{2}+\left|\Delta_{\alpha}({\bf k}) \right|^{2}} $.
Obviously, the superconductivity in ferromagnetic superconductors is
{\bf non-unitary}.

The disregard by the electromagnetic field $ 4\pi M$ acting on
electron charges does not mean the absence of  magnetic moment M due
to difference in the electron spin up and spin down populations.
For
the normal state moment density we obtain from the eqn (\ref{e38})
\begin{equation}
{\bf M}_{n}¥=\hat z \mu_{B}¥T\sum_{n}¥\int d\xi\left [
\frac{\langle \nu_{\uparrow}¥(\xi,\hat {\bf
p})\rangle}{i\omega_{n}¥-\xi}-\frac{ \langle \nu_{\downarrow
}¥(\xi,\hat {\bf p})\rangle}{i\omega_{n}¥-\xi}\right ]=\hat z
\mu_{B}¥( N_{\uparrow}¥-N_{\downarrow}¥),
\label{e42}
\end{equation}
where $\langle \nu_{\alpha}¥(\xi,\hat {\bf p})\rangle$ and
$N_{\alpha}¥$ are  the density of states averaged over the solid angle
and the
density of particles in the corresponding band.  The magnetic moment
in superconducting state acquires an extra value.  Near the critical
temperature it is
\begin{eqnarray}
{\bf M}_{s}¥&=&{\bf M}-{\bf M}_{n}¥=\hat z \mu_{B}¥T\sum_{n}¥\int
\xi d\xi\left [ \frac{\langle \nu_{\uparrow}¥(\xi,\hat {\bf
p})|\Delta_{\uparrow}¥({\bf p})|^{2}¥\rangle
}{(\omega_{n}¥^{2}¥+\xi^{2}¥)^{2}¥} - \frac{\langle \nu_{\downarrow
}¥(\xi,\hat {\bf p})|\Delta_{\downarrow }¥({\bf p})|^{2}¥\rangle }
{(\omega_{n}¥^{2}¥+\xi^{2}¥)^{2}¥}\right]\\
\nonumber &=&\frac{\hat z
\mu_{B}¥}{4}\left( \langle \nu_{\uparrow}¥'(0,\hat {\bf
p})|\Delta_{\uparrow}¥({\bf p})|^{2}¥\rangle \ln\frac{2\gamma\tilde
\epsilon_{\uparrow}¥}{\pi T_{c}¥}- \langle \nu_{\downarrow}¥'(0,\hat
{\bf p})|\Delta_{\downarrow}¥({\bf p})|^{2}¥\rangle
\ln\frac{2\gamma\tilde \epsilon_{\downarrow}¥}{\pi T_{c}¥}\right),
\label{e43}
\end{eqnarray}
where $\nu_{\alpha}¥'(0,{\bf p})$ is the derivative of the density of
states at the Fermi surface and $\tilde{\varepsilon}_{\alpha}¥$ is the pairing
interaction energy cutoff in the corresponding band.  It is
instructive to compare this expression with the "Cooper pair spin
momentum" (\ref{e27}).

\subsection{The order parameter equations near the superconducting 
transition}

We shall be interested by the simplest applications of the general
theory formulated in the previous subsection such that the critical
temperature suppression by the impurities and upper critical field
calculation .  For this purpose we need the system of the linear
equations for the order parameter arising from Gor'kov equations
averaged over impurities. This system consists of two equations for
the order parameter components with spin polarizations "up" and
"down",

\begin{eqnarray}
\Delta_{\alpha}\left({\bf R},{\bf r}\right)&=& -T \sum_{n,\beta}
\int d {\bf r}' V_{\alpha, \beta}\left({\bf r},{\bf r}'\right)
G^{0}¥_{\beta}¥({\bf r}',\tilde{\omega}_{n}¥^{\beta}¥) G^{0}¥_{\beta}¥({\bf
r}',-\tilde{\omega}_{n}¥^{\beta}¥) \nonumber \\
&&\times
\exp
\left[
i {\bf r}' {\bf D}({\bf R})
\right]
\left\{
\Delta_{\beta} \left({\bf R},{\bf r}'\right) +
\Sigma_{\beta}(\tilde{\omega}_{n}^{\beta},{\bf R}) \right\},
\label{e44}
\end{eqnarray}
and two equations for the impurity self-energy components
\begin{eqnarray}
\Sigma_{\alpha}\left(\tilde{\omega}_{n}^{\alpha},{\bf R}\right)&=&
n_{\mathrm{i}} u^{2}_{\alpha} \int d {\bf r}\, G^{0}¥_{\alpha}¥({\bf
r},\tilde{\omega}_{n}¥^{\alpha}¥) G^{0}¥_{\alpha}¥({\bf
r},-\tilde{\omega}_{n}¥^{\alpha}¥) \exp\left[ i {\bf r} {\bf D}({\bf
R}) \right] \nonumber \\
&&\times
\left\{
\Delta_{\alpha}\left({\bf R},{\bf r}\right) +
\Sigma_{\alpha}(\tilde{\omega}_{n}^{\alpha},{\bf R}) \right\}
\label{e45},
\end{eqnarray}
where
 $\tilde{\omega}_{n}^{\alpha}=\omega_{n}+{\mathrm sign} \, \omega_{n}/2
 \tau_{\alpha}$, and $\tau_{\alpha}$ is the quasi-particle mean free time in
 the different bands.
These mean free times are related in the Born approximation to the 
impurity concentration $n_{\mathrm{i}}$ through
\begin{equation}
\frac{1}{2 \tau_{\alpha}}= \pi n_{\mathrm{i}} N_{0\alpha}u^{2}_{\alpha} ,
\label{e46}
\end{equation}
with $u_{\alpha}$ - the amplitude of the impurity scattering and
$N_{0\alpha}$ - the density of electronic states in each band.

The operator of covariant differentiation is
$${\bf D}({\bf
R})=-i\frac{\partial}{\partial {\bf R}}+\frac{2e}{c}
{\bf A}({\bf R}).$$
The normal metal electron Green functions
are
\begin{equation}
G^{0}¥_{\alpha}¥({\bf r},\tilde{\omega}_{n}¥^{\alpha}¥) =\int \!  \frac{d
{\bf p}}{(2\pi)^{3}} e^{i {\bf p} \cdot {\bf r}}
\left(i\tilde{\omega}_{n}^{\alpha}-\xi_{{\bf p},\alpha}
+\mu_{B}{\bf g}_{{\bf p},\alpha}
{\bf H}/2\right)^{-1}.
\label{e47}
\end{equation}
The order parameter components in different bands are determined in
accordance with (\ref{e15}):
\begin{equation}
\Delta_{\uparrow}¥({\bf R},{\bf r})=
-\eta_{1}¥({\bf R})f_{-}¥(\hat {\bf r}),~~~~ \Delta_{\downarrow}¥({\bf
R},{\bf r})= \eta_{2}¥({\bf R})f_{+}¥(\hat {\bf r}).
\label{e48}
\end{equation}

Let us take now ${\bf H}_{ext}¥=0,  ~~~\nabla\times{\bf A}=4\pi{\bf M}$.
Consider two superconducting states $A_{1}¥$ and $A_{2}¥$ determined
by the equal functions $u_{i}¥^{A_{1}¥}¥=u_{i}¥^{A_{2}¥}¥$.  It is clear 
in this case that 
\begin{equation}
\Delta_{\uparrow}¥= -\eta_{1}¥f_{-}¥^{A_{1}¥}¥=-i\eta_{1}¥(f_{+}¥^{A_{2}¥}¥)^{*}¥,
\label{e49}
\end{equation}
\begin{equation}
\Delta_{\downarrow}¥=
\eta_{2}¥f_{+}¥^{A_{1}¥}¥=i\eta_{2}¥(f_{-}¥^{A_{2}¥}¥)^{*}¥.
\label{e50}
\end{equation}
We see, that if the superconducting state $A_{1}¥$ is a solution of the equations
(\ref{e44}), (\ref{e45}), then the state $A_{2}¥$ also obeys the same
equations but with interchanged band indices $1\leftrightarrow 2$ that
is accompanied according to eqns (\ref{e42}), (\ref{e43}) 
by the change of magnetic moment
direction. It means, that these $A_{1}¥$ and $A_{2}¥$ superconducting
states posess the same critical temperature but they are realized in
the ferromagnetic domains with the oppositely directed magnetic
moments.

\subsection{The critical temperature $T_{c0}$}

In the absence of an external field let us first find the critical temperature
$T_{c0}¥$ in the formally spacially uniform situation of negligible
electromagnetic field $4\pi{\bf M}=0$ acting diamagnetically on the
electron charges.  This case the anomalous impurity self-energy part
$\Sigma_{\alpha}(\widetilde{\omega}_{n}^{\alpha},{\bf R})=0$ and from
(\ref{e44}) we obtain the system of equations
\begin{eqnarray}
\eta_{1}¥ &=&(g_{1}¥\eta_{1}¥+
g_{12}¥\eta_{2}¥)\lambda(T_{c0}¥),\nonumber\\
\eta_{2}¥&=&(g_{21}¥\eta_{1}¥+
g_{2}¥\eta_{2}¥)\lambda(T_{c0}¥),
\label{e49}
\end{eqnarray}
where 
$g_{1}¥=
V_{\uparrow \uparrow}¥\langle|f_{-}¥({\bf k})|^{2}¥ N_{0\uparrow }¥
(\hat{\bf k})\rangle $, the angular brackets mean the
averaging over the Fermi surface, $N_{0\uparrow }¥(\hat{\bf k})$ is
the angular dependent density of electronic states at the Fermi
surface of the band $\uparrow$.  Correspondingly
$g_{12}¥=
V_{\uparrow \downarrow}¥\langle|f_{+}¥({\bf k})|^{2}¥ N_{0\downarrow }¥
(\hat{\bf k})\rangle $, $~~g_{21}¥=
V_{\downarrow\uparrow
}¥\langle|f_{-}¥({\bf k})|^{2}¥ N_{0\uparrow }¥ (\hat{\bf k})\rangle 
$, $~~g_{2}¥=V_{\downarrow \downarrow}¥\langle|f_{+}¥({\bf k})|^{2}¥ N_{0\downarrow }¥(\hat{\bf
k})\rangle $. 
The function $\lambda(T)$ is 
\begin{equation}
\lambda(T)=2\pi T\sum_{n\ge 0}¥\frac{1}{\omega_{n}¥}=\ln
\frac{2\gamma \epsilon}{\pi T}~,
\label{e50}
\end{equation}
$\ln\gamma=0,577\ldots $ is the Euler constant, $\epsilon $ is an
energy cutoff for the pairing interaction.  We assume here  that it has the
same value for both bands.
 
Thus, similar to \cite{30,31} the critical temperature is given by
\begin{equation}
T_{c0}¥=(2\gamma\epsilon/\pi)\exp{(-1/g)},
\label{e51}
\end{equation}
where $g$ is defined by the maximum of zeros of determinant of the system
(\ref{e49})
\begin{equation}
g= (g_{1}¥+g_{2}¥)/{2} +\sqrt{({g_{1}¥-g_{2}¥})^{2}¥/{4}+g_{12}¥g_{21}¥}.
\label{e52}
\end{equation}
In particular at $g_{12}¥,~g_{21}¥~\ll~g_{1}¥,~g_{2}¥$ the critical
temperature is determined by
\begin{equation}
g= \max(g_{1}¥,~g_{2}¥).
\label{e53}
\end{equation}

All the properties of metal depend on pressure.  In ferromagnetic
metal the pressure shifts the Fermi surface position and changes the
densities of spin up and spin down electron populations.  Density of
states at the Fermi suface in each band and the superconducting
interaction are also changed.  The critical temperature changes
following to the relative changes of the effective constant of pairing
interaction $g$.  In the case of one band pairing the latter can be
roughly represented as sum of relative change of density of states due
to the Fermi energy shift and relative change of pairing amplitude
${\delta g}/{g}\propto {\delta
\varepsilon_{F}¥}/{\varepsilon_{F}¥}+{\delta V}/{V}$.  The situation
with changes of the pairing interaction is far to be clear.  In
assumption that the relative change of density of states gives the
main contribution we have
\begin{equation}
T_{c}¥(P)=T_{c}¥(P_{0}¥)\left(1 +\frac{\delta \varepsilon_{F}¥} 
{\varepsilon_{F}¥}\ln \frac{2\gamma \epsilon}{\pi T_{c}¥(P_{0}¥)}\right).
\label{e54}
\end{equation}
The Fermi energy shift can be somehow magnetization dependent.  In
the simplest case one can expect $\delta\varepsilon_{F}¥\propto
\mu_{B}¥\delta M$.  Thus the magnetization changes can cause the
growing up (as well as falling down) of the superconducting transition
temperature.  This has been proposed \cite{25} as an explanation of a
{\bf "stimulation" of superconductivity by ferromagnetism} observed in
$ZrZn_{2}¥$ \cite{3}.  On the other hand a superconductivity is always
suppressed by the diamagnetic currents.  We shall make the comparison
of these two mechanisms of $T_{c}¥(P)$ dependence after the
calculation of the upper critical field.

\subsection{The critical temperature dependence on impurities concentration}

Triplet superconductivity is
suppressed by non-magnetic impurities \cite{55}.  The law of
suppression of superconductivity is given by the universal
Abrikosov-Gor'kov (AG) dependence \cite{56}
\begin{equation}
    -\ln t=\Psi\left( \frac{1}{2}+ \frac{x}{4\gamma t}\right)
-\Psi\left( \frac{1}{2}
\right)
\label{e55}
\end{equation} 
valid for any unconventional superconducting state and applicable 
in particular to a concrete unconventional superconductor 
independently of the
pressure \cite{24}. Here $\Psi$ is the digamma function. The 
variable $t=T_{c}•/T_{c0}•$ is the ratio
of the critical temperature of the superconductor with a given
concentration of impurities $n_{\mathrm{i}}$ to the critical 
temperature of the clean
superconductor, and $x 
=n_{\mathrm{i}}/n_{\mathrm{ic}}•=\tau_{\mathrm{c}}•/\tau$ is the 
ratio 
of the
impurity concentration in the superconductor to the critical 
impurity concentration destroying
superconductivity, or the inverse ratio of the corresponding mean 
free particle lifetimes.  The critical mean free time is given by 
$\tau_{\mathrm{c}}•=\gamma/\pi T_{c0}• $.  This dependence has been
demonstrated (although with some dispersion of the experimental
points) for the triplet superconductor
$\mathrm{Sr}_{2}•\mathrm{RuO}_{4}•$ \cite{57}.
    
Deviations from the universality of the
AG law can be caused by the anisotropy of the scattering which
takes place in the presence of extended imperfections in the 
crystal.  Such
a modification of the theory applied to
$\mathrm{UPt}_{3}•$ has been considered in the paper \cite{58}.  However, a
complete experimental investigation of the suppression of
superconductivity by impurities in this unconventional superconductor,
in particular the study of the universality of the behavior, has not
been performed.

The nonuniversality of the suppression of
superconductivity can also be caused by any inelastic   
scattering mechanism by impurities with internal degrees of
freedom of magnetic or nonmagnetic origin.  For the simplest
discussion of this, see \cite{59}. 

Finally, universality
is certainly not expected in multiband superconductors.  Theories 
for this case have been developed with regard to the unconventional
superconductivity in $\mathrm{Sr}_{2}•\mathrm{RuO}_{4}•$ (p-wave, 
two-band
two-dimensional model \cite{60}) and conventional 
superconductivity in
$\mathrm{MgB}_{2}•$ (anisotropic scattering two-band model 
\cite{61}).
    
A simple modification of the universal AG law
for the suppression of the superconductivity by impurities in a
two-band ferromagnetic superconductor is derived here.  Our consideration
is limited to the simplest case of scattering by ordinary point-like
impurities.  Then, due to spin conservation, one can neglect interband
quasi-particle scattering and take into account only the intraband
quasi-particle scattering on impurities.  
At finite impurity concentration the similar to (\ref{e49}) system of
equations is:
\begin{eqnarray}
\eta_{1}¥&=&g_{1}¥\Lambda_{1}¥(T)\eta_{1}¥+
g_{12}¥\Lambda_{2}¥(T)\eta_{2}¥,\nonumber\\
\eta_{2}¥&=&g_{21}¥\Lambda_{1}¥(T)\eta_{1}¥+ g_{2}¥\Lambda_{2}¥(T)\eta_{2}¥,
\label{e56}
\end{eqnarray}
where
\begin{equation}
\Lambda_{1,2}¥(T)= \Psi\left(\frac{1}{2}\right)-\Psi\left(\frac{1}{2}+
\frac{1}{4\pi\tau_{1,2}¥T}\right)+\ln\frac{T_{c0}¥}{T}+
\lambda(T_{c0}¥).
\label{e57}
\end{equation}

Hence the critical temperature is determined from the equation
\begin{equation}
(g_{1}¥\Lambda_{1}¥(T)-1) (g_{2}¥\Lambda_{2}¥(T)-1)-
g_{12}¥g_{21}¥\Lambda_{1}¥(T)\Lambda_{2}¥(T)=0.
\label{e58}
\end{equation}
In particular at $g_{12}¥,~g_{21}¥~\ll~g_{1}¥,~g_{2}¥$ the critical temperature
is determined by the $\max(T_{c1}¥,T_{c2}¥)$ of the solutions of equations
\begin{equation}
\ln\frac{T_{c0}¥}{T_{c1}¥}=\Psi\left(\frac{1}{2}+
\frac{1}{4\pi\tau_{1}¥T_{c1}¥}\right)-\Psi\left(\frac{1}{2}\right)+
\frac{1}{g_{1}¥}-\lambda(T_{c0}¥),
\label{e59}
\end{equation}
\begin{equation}
\ln\frac{T_{c0}¥}{T_{c2}¥}=\Psi\left(\frac{1}{2}+
\frac{1}{4\pi\tau_{2}¥T_{c2}¥}\right)-\Psi\left(\frac{1}{2}\right)+
\frac{1}{g_{2}¥}-\lambda(T_{c0}¥).
\label{e60}
\end{equation}
Let us accept for determiness that $g_{1}¥> g_{2}¥$ hence the  maximal
critical temperature in absence of impurities is defined by
$1/g_{1}¥=\lambda(T_{c0}¥)$.  Then at small impurity concentrations the
solutions of (\ref{e59}) and (\ref{e60}) are the linear functions of impurities
concentration:
\begin{equation}
T_{c1}¥=T_{c0}¥~-~\frac{\pi}{8\tau_{1}¥},
\label{e61}
\end{equation}
\begin{equation}
T_{c2}¥=T_{c0}¥-\frac{1}{g_{2}¥}+\frac{1}{g_{1}¥}~-~\frac{\pi}{8\tau_{2}¥}.
\label{e62}
\end{equation}
These lines can in principle intersect each other, as result an upturn
on the critical temperature dependence of impurity concentration
$T_{c}¥(n_{i}¥)$ is appeared.
Such the type of  deviations of the $T_{c}¥(n_{i}¥)$ dependence from
the AG-law present the direct manifestation of the two-band character of the
superconductivity.  On the other hand, an absence of strong deviations
from the universal one-band curve if it would found experimentally in
a ferromagnetic superconductor means that the superconductivity is
developed in one-band with only electrons with "up" spins paired and
the "down" spin electrons leave normal (or vice versa).

Another specific feature of the ferromagnetic superconductors is that even
in the absence of an external magnetic field
a ferromagnet produces an electromagnetic field $4\pi M \sim 4 \pi
{\mu_{ B}•k_{F}•^{3}•} $ acting via the electronic charges on the
orbital motion of electrons, and suppressing the
superconductivity~\cite{footnote}.  Hence, the actual critical
temperature in ferromagnetic superconductors is always smaller by the
value $\sim 4 \pi M •/H_{c2}•(T=0)$ relative to the (imaginary)
ferromagnetic superconductor without $4 \pi M•$.  The upper critical
field $H_{c2}•$ is also purity dependent.  That is why the impurity
concentration dependence of the actual $T_{c}•$ in a ferromagnetic
superconductor might be determined not only directly by the
suppression of superconducting correlations by the impurity scattering
as in any nonconventional superconductor but also indirectly through
the supression of $H_{c2}•$.  In fact the second indirect mechanism
has a smaller influence because the ratio $4 \pi M•/H_{c2}•(T=0)$ is
less than $1/10$ for superconductors with an upper critical field of
the order of several Teslas.

Thus the problem of determination of the critical temperature in
superconducting ferromagnet is at bottom the problem of determination
of the upper critical field in single domain ferromagnet.

\subsection{The upper critical field}

The equations for determination of upper critical field at least near
$T_{c}¥$ is easily derived from the system (\ref{e44}), (\ref{e45}). 
Keeping only the lowest order gradient terms we have
\begin{eqnarray}
\Delta_{\alpha}\left({\bf R},{\bf r}\right)&=& -T \sum_{n,\beta}
\int d {\bf r}' V_{\alpha, \beta}\left({\bf r},{\bf r}'\right)
G^{0}¥_{\beta}¥({\bf r}',\tilde{\omega}_{n}¥^{\beta}¥) G^{0}¥_{\beta}¥({\bf
r}',-\tilde{\omega}_{n}¥^{\beta}¥) \nonumber \\
&&\times
\left\{\left(1- \left({\bf r}' {\bf D}({\bf R}) \right)^{2}¥/2\right) 
\Delta_{\beta} \left({\bf R},{\bf r}'\right) +
\left(i{\bf r}' {\bf D}({\bf R}) \right)
\Sigma_{\beta}(\tilde{\omega}_{n}^{\beta},{\bf R}) \right\},
\label{e63}
\end{eqnarray}
and 
\begin{eqnarray}
\Sigma_{\alpha}\left(\tilde{\omega}_{n}^{\alpha},{\bf R}\right)&=&
n_{\mathrm{i}} u^{2}_{\alpha} \int d {\bf r}\, G^{0}¥_{\alpha}¥({\bf
r},\tilde{\omega}_{n}¥^{\alpha}¥) G^{0}¥_{\alpha}¥({\bf
r},-\tilde{\omega}_{n}¥^{\alpha}¥) \nonumber \\
&&\times\left\{\left( i {\bf r} {\bf D}({\bf R}) \right)
\Delta_{\alpha}\left({\bf R},{\bf r}\right) +
\Sigma_{\alpha}(\tilde{\omega}_{n}^{\alpha},{\bf R}) \right\}
\label{e64},
\end{eqnarray}
Finding $\Sigma_{\alpha}(\tilde{\omega}_{n}^{\alpha},{\bf R})$ from
the last equation and substituting to (\ref{e31}) we obtain after all
the necessary integrations the pair of the Ginzburg-Landau equations
for two components of the order parameter
\begin{eqnarray}
\eta_{1}¥&=&V_{\uparrow\uparrow}¥\hat\alpha_{1}¥\eta_{1}¥+
V_{\uparrow\downarrow}¥\hat\alpha_{2}\eta_{2}¥,\nonumber\\
\eta_{2}¥&=&V_{\uparrow\downarrow}¥\hat\alpha_{1}\eta_{1}¥+
V_{\downarrow\downarrow}¥\hat\alpha_{2}\eta_{2}¥,
\label{e65}
\end{eqnarray}
where operator $\hat\alpha_{1}$ consists of previously determined homogeneous part and 
second order gradient terms
\begin{equation}
\hat\alpha_{1}¥=\langle|f_{-}¥({\bf k})|^{2}¥ N_{0\uparrow }¥
(\hat{\bf k})\rangle\Lambda_{1}¥(T)-K_{\uparrow ij}¥D_{i}¥D_{j}¥.
\label{e66}
\end{equation}
The gradient terms coefficients are
\begin{eqnarray}
K_{\uparrow ij}¥&=&\langle|f_{-}¥({\bf k})|^{2}¥ N_{0\uparrow }¥ (\hat{\bf
k})v_{F\uparrow i}¥(\hat{\bf k})v_{F\uparrow j}¥(\hat{\bf k})\rangle
\frac{\pi T}{2}\sum_{n\ge
0}¥\frac{1}{|\tilde\omega^{\uparrow}¥_{n}¥|^{3}¥}\nonumber\\
&+&\langle
f_{-}¥({\bf k}) N_{0\uparrow }¥ (\hat{\bf k})v_{F\uparrow i}¥(\hat{\bf
k})\rangle\langle f_{-}¥^{*}¥({\bf k}) N_{0\uparrow }¥ (\hat{\bf
k})v_{F\uparrow j}¥(\hat{\bf k})\rangle \frac{\pi^{2}¥T
n_{i}¥u_{\uparrow}¥^{2}¥}{2} \sum_{n\ge
0}¥\frac{1}{\omega_{n}¥^{2}¥\tilde\omega^{\uparrow}¥_{n}¥^{2}¥}
\label{e67}
\end{eqnarray}
Operator $\hat\alpha_{2}$ is obtained from here by the natural
substitutions $1\to 2$, $\uparrow~ \to~ \downarrow$, $+~\to ~- $.

Now the problem of the upper critical field finding is just the
problem of solution of the two coupled equations (\ref{e65}).  There are a
lot of different situations depending of crystal symmetry, direction
of spontaneous magnetization and the external field orientation.  The
simplest case is when the external magnetic field is parallel or
antiparallel to the easy magnetization axis.  If the latter coinsides
with 4-th order symmetry axis in the cubic crystal like it is in
$ZrZn_{2}¥$ when the gradient terms in the perpendicular plane are
isotropic and described by two constants $K_{\uparrow
ij}¥=K_{\uparrow}¥\delta_{ij}¥$, $K_{\downarrow ij}¥=
K_{\downarrow}¥\delta_{ij}¥$.  This case formally corresponds to the
problem of determination of upper critical field parallel to the
c-direction in two-band hexagonal superconductor $MgB_{2}¥$ solved in
\cite{31}.  Then the linearized Ginzburg-Landau equations describe a
system of two coupled oscillators and have the solution in the form
$\eta_{1}¥=c_{1}¥f_{0}¥(x)$ and $\eta_{2}¥=c_{2}¥f_{0}¥(x)$, where
$f_{0}¥(x)=\exp(-hx^{2}¥/2)$ and $h$ is related to the upper critical field 
by means
\begin{equation} 
|H_{c2}¥\pm 4\pi M|=\frac{h(\tau)\Phi_{0}¥}{2\pi},
\label{e68}
\end{equation}
where $\Phi_{0}¥$ is the flux quantum. 

Let us for the simplicity limit
ourself by the impurityless case.  Then $\tau=1-T/T_{c0}¥$ and the
equation for the determination of upper critical field is
\begin{eqnarray}
&&\left[g_{1}¥(\tau+
\lambda(T_{c0}¥))+V_{\uparrow\uparrow}¥K_{\uparrow}¥h-1\right]
\left[g_{2}¥(\tau
+\lambda(T_{c0}¥))+V_{\downarrow\downarrow}¥K_{\downarrow}¥h-1\right]\nonumber\\
&-&
\left[g_{12}¥(\tau
+\lambda(T_{c0}¥))+V_{\uparrow\downarrow}¥K_{\uparrow}¥h-1\right]
\left[g_{21}¥(\tau
+\lambda(T_{c0}¥))+V_{\uparrow\downarrow}¥K_{\downarrow}¥h-1\right]=0
\label{e69}
\end{eqnarray}
This is a simple square equation and as before if we consider the case
$g_{12}¥, g_{21}¥\ll g_{1}¥, g_{2}¥$ and $g_{1}¥>g_{2}¥$ then we
obtain the following two roots
\begin{equation}
h_{1}¥(\tau)=\frac{g_{1}¥\tau}{V_{\uparrow\uparrow}¥K_{\uparrow}¥},
\label{e70}
\end{equation}
\begin{equation}
h_{2}¥(\tau)=\frac{g_{2}¥}{V_{\downarrow\downarrow}¥K_{\downarrow}¥}\left(
\tau+\frac{1}{g_{1}¥}-\frac{1}{g_{2}¥}\right)
\label{e71}
\end{equation}
This two lines can in principle intersect each other, then an upturn on the 
temperature dependence of the upper critical field given by the 
$max(h_{1}¥,h_{2}¥)$ is
appeared.

In the more anisotropic situation such as in orthorombic crystals $UGe_{2}¥$
and $URhGe$ even for the external field direction parallel or
antiparallel to the easy magnetization axis all the coefficients
$K_{\uparrow xx}$, $K_{\uparrow yy}¥$, $K_{\downarrow xx}¥$,
$K_{\downarrow yy}¥$ are different.  Then our system of equations can be
solved following a variational approach developed in \cite{31}.  Again
an upturn in $h(\tau)$ dependence can be possible.

The comparison with experiment shall be always not easy masked by the
presence of many ferromagnetic domains.  The monodomain measurements
are possible in high enough fields.  To work in this region one can
easily obtain the forth order gradient terms to the Ginzburg-Landau
equations.  However the problem of theoretical determination of the
upper critical field at arbitrary temperature has the same
principal difficulties as in any conventional anisotropic superconductor
\cite{63}.

At last, we shall discuss the problem of {\bf stimulation of
superconductivity by the ferromagnetism}.  The simple estimation from
the equation (\ref{e70}) shows that in the absence of an external
field the diamagnetic suppression of critical temperature of
ferromagnetic superconductor by its own ferromagnetic moment is
\begin{equation}
T_{c}¥\propto T_{c0}¥\left(1-\frac{\mu_{B}¥ M m v_{F}¥^{2}¥ }{T_{c0}¥^{2}¥
} \right).
\label{e72}
\end{equation}
Hence the comparison of this expression with formula (\ref{e54})
yields the criterium for the stimulation of supercondactivity by
ferromagnetism
\begin{equation}
\frac{\delta \varepsilon_{F}¥} 
{\varepsilon_{F}¥}~>~\mu_{B}¥ m ~\delta~\frac{ M
v_{F}¥^{2}¥}{T_{c0}¥^{2}¥}
\label{e73}
\end{equation}
It looks like unrealistic.  Hence the explanation of stimulation of
superconductivity by ferromagnetism in $ZrZn_{2}¥$ 
introduced in the paper \cite{25} seems unplausible.  We remind, however, that
criterium (\ref{e73}) was obtained in the assumption of absence of changes of
the pairing amplitude with pressure which could be the main source of
the critical temperature changes.

\section{Superconductivity in the metals without inversion centrum}

The multiband classification of electronic states in ferromagnetic metals
appeares as the result of the level splitting due to an exchange
field.  Another reason for the level splitting always exists in a
metal without inversion center.  This is spin-orbital coupling.  It
causes not only electron level splitting but also the nontrivial
spinor structure of the electronic states being important for the
proper description of superconductivity in the metals with broken
space parity.  The good introduction in general formalism of the Bloch
states and superconductivity in noncentrosymmetric crystals is given
in paper \cite{48} where also discussed the particular form of the
theory in the limit of small spin-orbit coupling.  Here in somethat
different manner we introduce the basic theoretical notions.  Unlike
to the paper \cite{48} and several others cited in Introduction we
shall speak only about the situation with strong spin-orbital coupling
producing the big band splitting and preventing the pairing of
electronic states from different bands.  The recently discovered
noncentrosymmetric superconductor $CePt_{3}¥Si$ belongs to this
cathegory.  The band structure calculation for this material found
\cite{47} that the bands for the states close to Fermi level are split
due to the spin-orbital coupling by 50-200 mev, which is more than
thousand times larger than the temperature of superconducting
transition $T_{c}¥=0.75K$ \cite{34}.  In this case the theory acquires
the features of similarity on the theory of superconductivity for
ferromagnetic superconductors.

\subsection{Electronic states and pairing in noncentrosymmetric metals}

Let us start from description of normal state in
the crystal without inversion centrum.  For each band its
single-electron Hamiltonian has the form
\begin{equation}
H=\varepsilon^{0}¥_{{\bf k}}¥+ \bbox{\alpha}_{{\bf k}}¥\bbox{\sigma},
\label{e74}
\end{equation}
where ${\bf k}$ is the wave-vector, the 
$\varepsilon^{0}¥_{{\bf k}}¥=\varepsilon^{0}¥_{-{\bf k}}¥$ is even function
of ${\bf k}$, the spin-orbital coupling is described
by an odd pseudovectorial function $\bbox{\alpha}_{{\bf k}}¥=
-\bbox{\alpha}_{-{\bf k}}¥$,
$\bbox{\sigma}=(\sigma_{x}¥,\sigma_{y}¥,\sigma_{z}¥)$ is the vector
consisting of Pauli matrices.  The eigen values and eigen functions of
this Hamiltonian are
\begin{equation}
\varepsilon_{{\bf k}\lambda }¥=\varepsilon^{0}¥_{{\bf k}}¥-\lambda
|\bbox{\alpha}_{{\bf k}}¥|,
\label{e75}
\end{equation}
\begin{eqnarray}
\Psi_{+ }¥({\bf k})=C_{\bf p}¥\left(
\begin{array}{c}
-\alpha_{{\bf k}x}¥+i\alpha_{{\bf k}y}¥ \\
\alpha_{{\bf k}z}¥+|\bbox{\alpha}_{{\bf k}}¥| \end{array}\right),~~~~~
\Psi_{-}¥({\bf k})=C_{\bf p}¥
\left(
\begin{array}{c}
\alpha_{{\bf k}z}¥+|\bbox{\alpha}_{{\bf k}}¥| \\
\alpha_{{\bf k}x}¥+i\alpha_{{\bf k}y}¥\end{array}\right),
\label{e76}
\end{eqnarray}
$$
C_{\bf p}¥=(2|\bbox{\alpha}_{{\bf k}}¥| (\alpha_{{\bf
k}z}¥+|\bbox{\alpha}_{{\bf k}}¥|))^{-1/2}¥.
$$
So, we have obtained the band splitting and $\lambda=\pm $ is the band index.
As result, there are two Fermi surfaces determined by equations
\begin{equation}
\varepsilon_{{\bf k}\lambda }¥=\varepsilon_{F}¥,
\label{e77}
\end{equation}
which may of course have the degeneracy points or lines for some
directions of ${\bf k}$.  The symmetry of directions of the dispersion laws
$\varepsilon_{{\bf k}\lambda }¥$ has to correspond to the crystal symmetry.
Particular attention however deserves the operation of reflection ${\bf k}$ to
$-{\bf k}$ which creates the time reversed states.

By application of operator of time
inversion $\hat K=-i\sigma_{y}¥K_{0}¥$, where $K_{0}¥$ is the
complex-conjugation operator one can see that the state $\Psi_{\lambda
}¥({\bf k})$ and the state inversed in time 
\begin{eqnarray}
\hat K \Psi_{\lambda
}¥({\bf k})&=&t_{\lambda}¥({\bf k})\Psi_{\lambda}¥(-{\bf k}),\nonumber \\
t_{\lambda}¥({\bf k})=e^{i\lambda(\pi +\phi_{\bf k}¥)}¥,~~~~~~~\phi_{\bf k}¥&=&\arg
\bbox{\alpha}_{{\bf k}\perp}¥, ~~~~ \bbox{\alpha}_{{\bf
k}\perp}¥=\alpha_{{\bf k}x}¥\hat x+\alpha_{{\bf k}y}¥\hat y
\label{e78}
\end{eqnarray}
are degenerate.  Another
words, they correspond to the same energy $\varepsilon_{{\bf k}\lambda
}¥=\varepsilon_{-{\bf k}\lambda }¥$.  So, the Fermi
surfaces in a crystal without inversion center still have mirror symmetry.
This is the consequence of time inversion symmetry.

Let us note also the important fact that the phase factor in (\ref{e78})
is the odd function of ${\bf k}$:
\begin{equation}
t_{\lambda}¥(-{\bf k})=-t_{\lambda}¥({\bf k}).
\label{e79}
\end{equation}

If we have the normal one-electron states classification in a crystal
without inversion symmetry it is quite natural to describe the
superconductivity directly in the basis of these states.  So, if we
consider the pairing only between the states with some ${\bf k}$ and
its negative then
the BCS Hamiltonian in the space homogeneous case looks as
follows
\begin{equation}
H_{BCS}¥=\sum_{{\bf k},\lambda}\xi_{{\bf k}\lambda }¥ a^{\dag}_{ {\bf
k}\lambda} a_{{\bf k}\lambda} +\frac{1}{2} \sum_{{\bf k},{\bf
k}',\lambda,\nu} V_{\lambda \nu}({\bf k},{\bf k}') a^{\dag}_{ -{\bf
k}, \lambda} a^{\dag}_{ {\bf k},\lambda} a_{ {\bf k}', \nu} a_{ -{\bf
k}', \nu},
\label{e80}
\end{equation}
where 
$\lambda, \nu=\pm$ are the band indices for the bands intoduced above
and 
\begin{equation}
\xi_{{\bf k}\lambda }¥=\varepsilon_{{\bf k}\lambda }¥-\mu
\label{e81}
\end{equation}
are the
band energies counted from the chemical potential.  Due to big
difference between the Fermi momenta we neglect in Hamiltonian by the
pairing of electronic states from different bands.  The structure of
theory is now very similar to the theory of ferromagnetic
superconductors with triplet pairing.  However, here there is some
special pecularity. The operators $a_{ {\bf k},\lambda}$ and $a^{\dag}_{
{\bf k},\lambda}$ with fixed band index $\lambda$, that are related to
the states in one particular band, they still are the spinor operators.
In particular, the time inversion transforms them in accordance with the rule
(\ref{e78}): 
\begin{equation}
Ka^{\dag}_{ {\bf k}, \lambda}=t_{\lambda}¥({\bf k})a^{\dag}_{
-{\bf k}, \lambda}, ~~~~~~~~~Ka_{ {\bf k},
\lambda}=t_{\lambda}¥^{*}¥({\bf k})a_{ -{\bf k}, \lambda}.
\label{e82}
\end{equation}

Let us introduce as usual the molecular fields
\begin{equation}
\Delta_{{\bf k}\lambda}=-\sum_{{\bf k}'}\sum_{\nu}
V_{\lambda\nu}\left( {\bf k},{\bf k}'\right) \langle a_{ {\bf k}',
\nu} a_{ -{\bf k}', \nu}\rangle .
\label{e83}
\end{equation}
Then the hamiltonian can be rewritten as
\begin{equation}
H_{BCS}¥=\sum_{{\bf k},\lambda}\xi_{{\bf k}\lambda }¥ a^{\dag}_{ {\bf
k}\lambda} a_{{\bf k}\lambda} +
\frac{1}{2} \sum_{{\bf k},\lambda} \Delta_{{\bf k}\lambda}
a^{\dag}_{ -{\bf k}, \lambda}a^{\dag}_{ {\bf k}, \lambda}+ \Delta_{{\bf k}\lambda}^{\dag}¥
a^{\dag}_{ {\bf k}, \lambda}a^{\dag}_{- {\bf k},\lambda}  ~~~+~~~const.
\label{e84}
\end{equation}
It follows immediately from the anticommutation of the Fermi operators
\cite{47} that
\begin{equation}
\Delta_{-{\bf k},\lambda}=-\Delta_{{\bf k},\lambda}.
\label{e85}
\end{equation}
On the other hand the hamiltonian (\ref{e84}) should be time reversal
invariant.  By application $\hat K$ to (\ref{e84}) and using rule
(\ref{e82}) and property (\ref{e79}) we find the condition of the time
invariance:
\begin{equation}
t_{\lambda}¥^{2}¥({\bf k})\hat K \Delta_{-{\bf k},\lambda}=
-\Delta_{{\bf k},\lambda}.
\label{e86}
\end{equation}
The solution of this equation is
\begin{equation}
\Delta_{{\bf k}\lambda}=t_{\lambda}¥({\bf k})\chi({\bf k}),
\label{e87}
\end{equation}
where $\chi({\bf k})$ is an even function of ${\bf k}$, which is easily
established from (\ref{e83}) if we chose for the pairing potential
\begin{equation}
V_{\lambda\nu}\left( {\bf k},{\bf k}'\right)=
V_{\lambda\nu}(k,k')t_{\lambda}¥({\bf k})t_{\nu}¥^{*}¥({\bf k}')
\sum_{i}^d\varphi_{i{\lambda}}¥({\bf k})\varphi_{i{\nu}}¥^{*}¥({\bf k}').
\label{e88}
\end{equation}
Here $\varphi_{i{\lambda}}¥({\bf k})$ are the even fuctions of an
irreducible representation dimensionality $d$ of the group of the
crystal symmetry $G$. 

Thus, in the noncentrosymmetric crystal the decomposition of the pairing
potential over the functions of irreducible representation contains
nontrivial phase factors.  The latter are odd functions of ${\bf k}$
and due to this reason the order parameter function $\Delta_{{\bf k},\lambda}$
being according to (\ref{e85}) an odd function of ${\bf k}$ is
transforming at the same time according to even function of
irreducible representation of the crystal symmetry group
\begin{equation}
\Delta_{{\bf k},\lambda}\propto t_{\lambda}¥({\bf k})\phi_{i{\lambda}}¥({\bf k}) .
\label{e89}
\end{equation}

The group of symmetry of the compound
$CePt_{3}¥Si$ is $C_{4v}¥$.  It has four one-dimensional irreducible
representations: $A_{1}¥, A_{2}¥, B_{1}¥, B_{2}¥$ and one
two-dimensional $E$.  The examples of even functions of its irreducible
representations are $\varphi_{A_{1}¥}¥\propto
(k_{x}^{2}¥+k_{y}^{2}¥+ck_{z}^{2}¥), ~~~\varphi_{A_{2}¥}¥\propto
k_{x}k_{y}(k_{x}^{2}¥-k_{y}^{2}¥), ~~~\varphi_{B_{1}¥}¥\propto
(k_{x}^{2}¥-k_{y}^{2}¥), ~~~\varphi_{B_{2}¥}¥\propto k_{x}k_{y},
~~~(\varphi_{E,1}¥,\varphi_{E,2}¥)\propto (k_{z}k_{x},k_{z}k_{y})$. 
For given superconducting state the functions $\varphi _{i\lambda}¥$
relating to the different bands  can be in principle different
functions transforming according to the same irreducible
representation.  The {\bf symmetry dictated nodes } in the
quasiparticle spectrum of superconducting $CePt_{3}¥Si$ are absent in
the case of realization of $A_{1}¥$ state.  They are placed on the
lines: $k_{x}¥=0, k_{y}¥=0, k_{x}¥=\pm k_{y}¥$ for $A_{2 }¥$ state,
$k_{x}¥=\pm k_{y}¥$ for $B_{1 }¥$ state, $k_{x}¥=0, k_{y}¥=0$ for
$B_{2}¥$ state, and $k_{z}¥=0$ and $k_{x}¥=k_{y}¥=0$ for $E$ state.

For {\bf Gor'kov equations in each band} we have
\begin{eqnarray}
&&\left(i\omega_{n}-\xi_{{\bf k}\lambda}\right) G_{\lambda}({\bf
k},\omega_{n})+ \Delta_{{\bf k}\lambda} F_{\lambda}^{\dagger}({\bf
k},\omega_{n})=1  \\
&&\left(i\omega_{n}+\xi_{-{\bf k}\lambda}\right) F_{\lambda}^{\dagger}¥ ({\bf
k},\omega_{n})+\Delta_{{\bf k}\lambda}^{\dagger}¥  G_{\lambda}({\bf
k},\omega_{n})=0.
\label{90}
\end{eqnarray}
The equations for each band are only coupled through the order
parameters given by the self-consistency equations
\begin{equation}
\Delta_{{\bf k}\lambda}=-T\sum_{n}\sum_{{\bf
k}'}\sum_{\nu} V_{\lambda\nu}\left( {\bf k},{\bf k}'\right)
F_{\nu}({\bf k}',\omega_{n}).
\label{e91}
\end{equation}
The superconductor Green's functions are
\begin{eqnarray}
G_{\lambda}\left({\bf k},\omega_{n}\right) &=&
\frac{i\omega_{n}+\xi_{-{\bf k}\lambda}}
{(i\omega_{n}-\xi_{{\bf k}\lambda})(i\omega_{n}+\xi_{-{\bf k}\lambda})
-\Delta_{{\bf k}\lambda}\Delta_{{\bf k}\lambda}^{\dagger}¥ } \\
F_{\lambda}\left({\bf k},\omega_{n}\right)&=& \frac{-\Delta_{{\bf k}\lambda} 
}{(i\omega_{n}-\xi_{{\bf k}\lambda})(i\omega_{n}+\xi_{-{\bf k}\lambda})
-\Delta_{{\bf k}\lambda}\Delta_{{\bf k}\lambda}^{\dagger}¥ }.
\label{e92}
\end{eqnarray}
The energies of elementary excitations are given by
\begin{equation}
E_{{\bf k}\lambda}¥=
\frac{\xi_{{\bf k}\lambda}¥-\xi_{-{\bf k}\lambda}¥}{2}\pm\sqrt
{\left(\frac{\xi_{{\bf k}\lambda}¥+\xi_{-{\bf k}\lambda}¥}{2}
\right)^{2}¥+
\Delta_{{\bf k}\lambda}\Delta_{{\bf k}\lambda}^{\dagger}¥ }.
\label{e93}
\end{equation}

The structure of the Gor'kov theory in the ferromagnetic and 
noncentrosymmetric superconductors has only formal similarity.
If in two band ferromagnets  the states in different bands have fixed
opposite spin projections, in two band noncentrocymmetric crystal the
states in each band are the spinors with spin projection depending on
momentum direction.  To make this distinction more transparent let us
write the {\bf Gor'kov equation in the initial spinor basis}, consisting of
two states with spin up and spin down projection:
\begin{eqnarray}
&&\left(i\omega_{n}-\xi^{0}¥_{{\bf k}}¥- \bbox{\alpha}_{\bf
k}\bbox{\sigma}\right)\hat G({\bf k},\omega_{n})+ \hat\Delta_{{\bf
k}} \hat F^{\dagger}({\bf k},\omega_{n})=\hat 1 \\
&&\left(i\omega_{n}+\xi^{0}¥_{-{\bf k}}¥+ \bbox{\alpha}_{-{\bf
k}}\bbox{\sigma}^{t}¥\right)\hat  F^{\dagger}¥ ({\bf
k},\omega_{n})+\hat \Delta_{{\bf k}}^{\dagger}¥\hat G({\bf
k},\omega_{n})=0,
\label{e94}
\end{eqnarray}
where
$\xi^{0}¥_{{\bf k}}¥=\varepsilon^{0}¥_{{\bf k}}¥-\mu$,
\begin{equation}
\hat G({\bf k},\omega_{n})=\hat {\bf  P}_{+}¥ G_{+}¥({\bf k},\omega_{n})+
\hat{\bf P}_{-}¥ G_{-}¥({\bf k},\omega_{n}),
\label{e95}
\end{equation}
\begin{equation}
\hat F^{\dagger}({\bf k},\omega_{n})=\hat g^{t}¥\left\{\hat {\bf  P}_{+}¥
F^{\dagger}_{+}¥ ({\bf k},\omega_{n})+\hat
{\bf P}_{-}¥F^{\dagger}_{-}¥({\bf k},\omega_{n})\right\},
\label{e96}
\end{equation}
\begin{equation}
\hat \Delta_{{\bf k}}=\left \{\hat {\bf P}_{+}¥\Delta_{{\bf k},+}+ {\bf
P}_{-}¥\Delta_{{\bf k},-}\right\}\hat g,
\label{e97}
\end{equation}
$\hat{\bf P}_{\pm }¥=(1\pm \hat {\bbox{\alpha}}_{\bf
k}\bbox{\sigma})/2$, $\hat g=i\hat \sigma_{y}¥$, $\hat
{\bbox{\alpha}}_{\bf k}= \bbox{\alpha}_{\bf k}/|\bbox{\alpha}_{\bf
k}|$.

Thus, superconducting order parameter consists of sum singlet and triplet
parts 
\begin{equation}
\hat \Delta_{{\bf k}}=
\frac{\Delta_{{\bf k},+}+\Delta_{{\bf k},-}}{2}\hat g+
\frac{\Delta_{{\bf k},+}-\Delta_{{\bf k},-}}{2}
\hat {\bbox{\alpha}}_{\bf k}\bbox{\sigma}\hat g.
\label{e98}
\end{equation}
At the same time in the absence of the external field the
superconducting state is unitary
\begin{equation}
{\bf M}=\mu_{B}¥T\sum_{n}¥\sum_{\bf k}¥\bbox{\sigma}
\hat G({\bf k},\omega_{n})=0.
\label{e99}
\end{equation}

It is worth noting that the basic equations (102)-(107) have the same
structure as in the theory with weak spin-orbital interaction
initially developed in \cite{35}.  There is however an important
distinction that the pairing potential is given now by eqn
(\ref{e88}).  On the other hand one can naively start from the pairing
of the "initial" states which are formed in the crystal with strong
spin- orbital coupling which do not introduce the parity violation and
only after this to add the parity violating terms.  Then, for the
triplet case with vector ${\bf d}({\bf k})$ of the order parameter,
the theory acquires very complicated form originating of presence of
the three physically different vectors ${\bf d}({\bf k})$,
$\bbox{\alpha}({\bf k})$ and ${\bf d}({\bf k}) \times
\bbox{\alpha}({\bf k})$.

\subsection{Electronic states in noncentrosymmetric metal and
pairing under magnetic field}

Let us look now on the modifications which are appeared by the application of
external magnetic field.   

It is known \cite{64} that the diamagnetic influence of field  is
taken into consideration by the Peierls substitution ${\bf k}\to{\bf
k}+(e/\hbar c){\bf A}(\partial/\partial{\bf k})$.  We shall be
interested here
in pure paramagnetic effects.  Neglecting by the term with magnetic field
in the Peierls substitution we take into account only direct
paramagnetic influence of magnetic field
\begin{equation}
H=\varepsilon^{0}¥_{{\bf k}}¥+ \bbox{\alpha}_{{\bf k}}¥\bbox{\sigma}
-\bbox{\mu}_{{\bf k}i}¥H_{i}¥\bbox{\sigma},
\label{e100}
\end{equation}
where $\bbox{\mu}_{{\bf k}i}¥=\bbox{\mu}_{-{\bf k}i}¥$ 
is even tensorial function of ${\bf k}$.  In the isotropic approximation
$\mu_{ij}¥=\mu_{B}¥g\delta_{ij}¥/2$, where $g$ is gyromagnetic ratio.
The eigen values of this Hamiltonian 
are
\begin{equation}
\varepsilon_{{\bf k}\lambda }¥=\varepsilon^{0}¥_{{\bf k}}¥-\lambda
|\bbox{\alpha}_{{\bf k}}¥-\bbox{\mu}_{{\bf k}i}¥H_{i}¥|.
\label{e101}
\end{equation}
It is obvious from here
that the time reversal symmetry is lost
$\varepsilon_{-{\bf k}\lambda }¥\ne\varepsilon_{{\bf k}\lambda }¥$ 
and the shape of the Fermi 
surfaces
do not obey the mirror symmetry.  The same situation takes place in the
{\bf ferromagnetic metal without inversion symmetry}.  The degeneracy of states
${\bf k}$ and$-{\bf k}$ is lifted by the exchange field and, in general,
in a ferromagnet without inversion like $MnSi$ can not be
superconducting.  On the contrary the discovery of superconductivity
in monoclinic ferromagnet $UIr$ is already reported \cite{65}.  It
could be either due cristalline anisotropy leading to weak influence
of exchange field on some group of charge carriers, or due to
realization of more exotic possibility like FFLO state.  The first
possibility is related to the problem of the paramagnetic limiting
field in noncentrosymmetric superconductors \cite{51} which we discuss
here.

For simplicity let us assume that we have pairing only in one band: 
$\lambda=+$.
The treatment of general case is similar but more lengthly.
Also we are limited ourselves by consideration only one-dimensional
representations when we have $V_{++}({\bf k},{\bf k}')=V~t({\bf
k})t^{*}¥({\bf k}')\varphi(\hat{\bf k})\varphi¥(\hat{\bf k}')$.  The
equation for critical temperature that is the linear version of
(\ref{e91}) has in this case the form
\begin{eqnarray}
\chi({\bf k})&=&-VT\sum_{n}\sum_{{\bf k}'} \varphi(\hat{\bf
k})\varphi^{*}¥(\hat{\bf k}') G^{0}¥({\bf k}',\omega_{n})\chi({\bf
k}')¥G^{0}¥(-{\bf k}',-\omega_{n}) \\ \nonumber &=&-VT\sum_{n}\sum_{{\bf
k}'} \frac{\varphi(\hat{\bf k})\varphi^{*}¥(\hat{\bf k}') \chi({\bf
k}')} {(i\omega_{n}¥-\xi_{{\bf k}'}¥)(-i\omega_{n}¥-\xi_{-{\bf k}'}¥)}.
\label{e102}
\end{eqnarray}
Here $\xi_{{\bf k}}¥=\varepsilon_{{\bf k}}-\mu$ and 
$\varepsilon_{{\bf k}}$ is given by (\ref{e101}).  We  also have taken into
consideration the relation (\ref{e87}).

It is  clear that the coherence between the normal metal states with
states with Green functions $G^{0}¥({\bf k},\omega_{n})$ and
$G^{0}¥(-{\bf k},-\omega_{n})$ is broken by magnetic field.  The
oppositely directed momenta ${\bf k}$ and $-{\bf k}$ on the Fermi
surface have the different length.  Hence the magnetic field will
suppress superconductivity that means the critical temperature will be
decreasing function of magnetic field.  It is clear also that it will
be anisotropic function of the field orientation in respect of
cristallographic directions.

For tetragonal crystal $CePt_{3}¥Si$ one can take as the simplest form
of gyromagnetic tensor $\mu_{ij}¥=\mu_{B}¥(g_{\perp}¥
(\hat x_{i}¥\hat x_{j}¥+\hat y_{i}¥ \hat y_{j}¥)+
g_{\parallel}¥\hat z_{i}¥\hat z_{j})/2$ and the pseudovector function
$\bbox{\alpha}_{{\bf k}}¥=\alpha(\hat z\times {\bf k})+ \beta\hat z
k_{x}¥k_{y}¥k_{z}¥(k_{x}¥^{2}¥-k_{y}¥^{2}¥)$. 
The latter is chosen following the discussion in the paper \cite{50}. 
Then for the normal metal energy of excitations we have
\begin{equation}
\xi_{\bf k}¥=\xi^{0}¥_{\bf k}¥-
\sqrt{(\alpha k_{y}¥+\frac{g_{\perp}¥}{2}\mu_{B}¥H_{x}¥)^{2}¥+
(\alpha k_{x}¥-\frac{g_{\perp}¥}{2}\mu_{B}¥H_{y}¥)^{2}¥+ 
(\beta k_{x}¥k_{y}¥k_{z}¥(k_{x}¥^{2}¥-k_{y}¥^{2}¥)
-\frac{g_{\parallel}¥}{2}\mu_{B}¥H_{z}¥)^{2}¥}
\label{e103}
\end{equation}

As result of simple calculation near $T_{c}¥$ we obtain
\begin{equation}
T_{c}¥({\bf H})=T_{c}¥\left\{1-
\frac{7\zeta(3)\mu_{B}¥}{32\pi^{2}¥T_{c}¥^{2}¥}
\left (a g_{\perp}¥^{2}¥(H_{x}¥^{2}¥+H_{y}¥^{2}¥)+
b g_{\parallel}¥^{2}¥H_{z}¥^{2}¥\right )+\ldots \right \},
\label{e104}
\end{equation}
that looks like similar to usual superconductivity with singlet pairing.
Here $a$ and $b$ are coefficients of the order of unity.  Its exact
values depend on the particular form of $\varphi(\hat{\bf k})$ functions
in pairing interaction as well on particular form of $\bbox{\alpha}_{{\bf k}}¥$.

On the other hand, let as assume that due to some particular reason 
coefficient $\beta$ is small.  Then for the field direction 
${\bf H}=H\hat z$ for $\mu_{B}¥g_{\parallel}¥H\gg\beta k_{F}¥^{5}¥$ we
have for the excitations energy
\begin{equation}
\xi_{\bf k}¥=\xi^{0}¥_{\bf k}¥-
\sqrt{(\alpha k_{y}¥)^{2}¥+ (\alpha
k_{x}¥)^{2}¥+ (\frac{g_{\parallel}¥}{2}\mu_{B}¥H_{z}¥)^{2}¥},
\label{e105}
\end{equation}
that is now the even function of the wave vector
$\xi_{\bf k}¥=\xi_{-{\bf k}}¥$.

The equation for the critical temperature has the form
\begin{equation}
\chi({\bf k})= -VT\sum_{n}\int d\xi N_{\xi=0}¥(\hat {\bf k}')
\frac{dS_{\hat {\bf k}'}¥}{S_{F}¥}\frac{\varphi(\hat{\bf
k})\varphi^{*}¥(\hat{\bf k}') \chi({\bf k}')} {(i\omega_{n}¥-\xi)
(-i\omega_{n}¥-\xi)}.
\label{e106}
\end{equation}
Here we can first integrate over the energy variable
$\xi$ and and then over the Fermi suface.  After the first integration
the magnetic field dependence is disappeared from equation and we
obtain standart BCS formula $T_{c}¥=(2\gamma/\pi)\epsilon\exp(-1/g)$
for critical temperature determination.  So, the suppression of critical
temperature by magnetic field is saturated at finite value which
differs from its value at $H=0$ due to field variation of density of
states and pairing interaction at $\xi=0$.

This results can be in principle valid for any direction of magnetic field
if paramagnetic interaction exceeds a spin-orbital splitting
$|\bbox{\mu}_{i}¥H_{i}¥|>|\bbox{\alpha}|$. 
Of course the superconductivity in the region of the
large fields still exists if $g$ is positive on the Fermi surface
$\xi=0$.  Thus at large fields the situation is similar to that we
have in the supercoductors with triplet pairing.

We have demonstrated that the paramagnetic suppression of superconducting
state in a crystal without inversion centrum certainly exists and the
effect depends of field orientation in respect of crystall axes.  The
paramagnetic suppression of superconductivity takes place due to
magnetic field lifting of degeneracy of electronic states with
opposite momenta ${\bf k}$ and $-{\bf k}$ forming the Cooper pairs. 
For some directions of fields the degeneracy is recreated.  That is
why the paramagnetic limit of superconductivity in the crystals
without inversion can be in principle absent.

To demonstrate the time inversion violation in its pure form we have
calculated the paramagnetic influence of external field on
superconductivity in the noncentrosymmetric material in complete
neglect of the diamagnetic currents.  Certainly the latter play the main
role in the superconductivity suppression.  The general Gor'kov
equations in this case have the same form (35)-(38) etc as for two band
ferromagnet.  One needs to remember only that the electron states and
energies in these two cases have quite different spinor structure and
parity in respect to ${\bf k}$.  One can find the treatment of several
inhomogeneous problems as the upper critical field calculation in
Ginzburg-Landau region in the papers \cite{40,49,50}.


\begin{thebibliography}{99}
\bibitem{footnote}
This is correct of course only if $4 \pi M > H_{c1}$ (where 
$H_{c1}$  is  the lower critical field).
\end{thebibliography}
\end{document}